\documentclass[conference]{IEEEtran}

 \usepackage{cite,url}
  \usepackage{amssymb,amsmath,amsfonts,amssymb,verbatim,graphicx,bm,color,amsbsy,amsthm,algorithm,algorithmic,epsfig,subfigure}
\usepackage{epstopdf}
\usepackage{caption}
\usepackage[top=0.75in, bottom=0.75in, left=0.75in, right=0.75in]{geometry}

\DeclareMathOperator*{\argmax}{argmax}

\begin{document}

%
\title{\vspace{0.25in} 
Noncoherent compressive channel estimation for mm-wave massive MIMO
}

\author{\IEEEauthorblockN{Maryam Eslami Rasekh\IEEEauthorrefmark{1}, Upamanyu Madhow\IEEEauthorrefmark{2}}
\IEEEauthorblockA{Department of Electrical and Computer Engineering\\
University of California Santa Barbara\\
Email: \IEEEauthorrefmark{1}rasekh@ucsb.edu, \IEEEauthorrefmark{2}madhow@ece.ucsb.edu}
}

\maketitle

\begin{abstract}
Millimeter (mm) wave massive MIMO has the potential for delivering orders of magnitude increases in mobile data rates, with compact antenna arrays 
providing narrow steerable beams for unprecedented levels of spatial reuse.  A fundamental technical bottleneck, however, is rapid spatial channel estimation and beam adaptation in the face
of mobility and blockage.  Recently proposed compressive techniques which exploit the sparsity of mm wave channels are a promising approach to this problem, with overhead scaling linearly with the number of dominant paths and logarithmically with the number of array elements.  Further, they can be implemented with RF beamforming with low-precision phase control. However, these methods make implicit assumptions on long-term phase coherence that are not satisfied by existing hardware. In this paper, we propose and evaluate a {\it noncoherent} compressive
channel estimation technique which can estimate a sparse spatial channel based on received signal strength (RSS) alone, and is compatible with off-the-shelf hardware.  
The approach is based on cascading phase retrieval (i.e., recovery of complex-valued measurements from RSS measurements, up to a scalar multiple)  
with coherent compressive estimation. While a conventional cascade scheme would multiply two measurement matrices to obtain an overall matrix whose entries
are in a continuum, a key novelty in our scheme is that we constrain the overall measurement matrix to be implementable using coarsely quantized pseudorandom phases, 
employing a {\it virtual} decomposition of the matrix into a product of measurement matrices for phase retrieval and compressive estimation.  Theoretical and simulation results show that our
noncoherent method scales almost as well with array size as its coherent counterpart, thus inheriting the scalability and low overhead of the latter.

\end{abstract}


\section{Introduction}

Emerging mm wave mobile networks have the potential to deliver several orders of magnitude increases in both per-user and network capacity, using a dense deployment of
small cell base stations and aggressive spatial reuse.  The small wavelengths allows scaling to massive MIMO with a very large number of elements within a form factor compatible
with compact base stations deployable, for example, on lampposts. The pencil beams enabled by such arrays enables drastically increased spatial reuse compared
to existing cellular networks. A critical challenge in realizing such mm wave mobile links, however, is the agile adaptation of such large arrays to track mobile users, while accounting for frequently occurring blockages.


As array sizes grow, simplification of front-ends is critical. Existing mm wave transceivers employ RF beamforming in place of costly per-element baseband control. This means a single I/Q stream is upconverted at the transmitter and distributed to all array elements, and the phase of each element is tuned with  RF phase shifters. At the receiver, the phase of each element is manipulated at RF and the combination of element outputs provides a single I/Q stream for signal processing. Thus, classical least squares techniques, which require individual access to the I/Q signal at each antenna element, cannot be applied for adaptive beamforming.


Conventional techniques for discovering spatial paths when constrained to RF beamforming include exhaustive and hierarchical scan.
In exhaustive scan, the transmitter
scans the entire angular domain with its narrow beam 
to identify the strongest path(s) to the user. The number of measurements  scales linearly with array size, hence the overhead becomes prohibitive for large arrays. In hierarchical
scan, the entire angular space is initially scanned with a small number of broad beams, with feedback from the receiver used to successively narrow the search space.
The number of measurements scales logarithmically with array size, but waiting for feedback from the receiver before each scan can be highly time-consuming and impractical upon implementation. This method does not scale well with the number of users, since each user may require a different beacon sequence, depending on its location and feedback. Neither of these methods is therefore suitable for low-overhead tracking for mm wave massive MIMO.

A promising alternative, which is our starting point in this paper, is to employ {\it compressive} techniques, introduced in \cite{ramasamy2012compressive,ramasamy_allerton12} and discussed in detail in the context of picocellular networks in \cite{marzi2016compressive}, which leverage the inherent sparsity of the mm wave channel to track users with a small number of measurements. In this approach, the transmitter broadcasts beacons using pseudorandom phases, and uses feedback from the receiver regarding the complex gain observed for each beacon to estimate the spatial channel by identifying the dominant paths. Such schemes can be implemented using RF beamforming (i.e., a single RF chain, rather than one RF chain per element) with severely quantized phase control, which allows simplification of the RF front end.  

We term the approach in  \cite{ramasamy2012compressive,ramasamy_allerton12,marzi2016compressive} {\it coherent} compressive estimation, since the receiver
must maintain phase coherence across successive measurements in order to provide the desired complex gains as feedback to the transmitter.
Unfortunately, such an approach does not work with commodity hardware, since current mm wave systems such as the 802.11ad standard are not
designed to maintain phase coherence across packets, and the oscillator offset and drift between the transmitter and receiver can alter the phase of each channel measurement randomly. This motivates us to develop the {\it noncoherent} compressive estimation approach presented in this paper.

\noindent
{\bf Contributions:} We use the same compressive beacon strategy as in prior work \cite{ramasamy2012compressive,ramasamy_allerton12,marzi2016compressive}, but provide an algorithm that can estimate a sparse spatial channel from RSS measurements alone. It does not require phase coherence across beacons, and can therefore be realized with commodity hardware. \\
1) We propose and evaluate a two-stage algorithm. The first stage is phase retrieval, in which the phase of the RSS measurements is recovered up to a constant phase offset. The second stage is
coherent compressive estimation on the the output of the phase retrieval stage.\\
2) The two-stage approach requires decomposition of the measurement matrix into a product of a phase retrieval matrix and a compressive measurement
matrix.  It is known that matrices with independent and identically distributed (i.i.d.) complex Gaussian entries are effective for this purpose, but the product of
such matrices has entries with complex values lying in a continuum, which cannot be realized with coarse phase control.  A key innovation in our proposed
approach is to constrain the {\it product} of the two matrices (i.e., the actual measurement matrix) to be implementable with coarse phase control, 
and to decompose it into two virtual matrices: an inner matrix that is used for coherent compressive estimation, and an outer matrix that is used for phase retrieval. 
That is, we choose one of the matrices, and infer the other via a pseudoinverse.  We provide design guidelines for the (non-unique) virtual decomposition,
and  demonstrate its performance and scalability through simulations.\\
3) Using a central limit theorem argument, we show that the number of measurements for sparse channel recovery with noncoherent compressive estimation scales only slightly worse than for coherent estimation.

\noindent {\bf Related work:} As mentioned, our starting point here is the work on coherent compressive estimation in \cite{ramasamy2012compressive,ramasamy_allerton12,marzi2016compressive}.
In our own prior work \cite{rasekh2017noncoherent}, noncoherent compressive channel estimation on the continuum is performed using noncoherent template matching and Newton refinement. This approach, while optimal for a single path channel, cannot be applied to a multipath channel with paths of comparable strength due to the nonlinear combination of noncoherent beacon responses.

In \cite{abari2016millimeter}, noncoherent tracking of multipath channels is attempted by designing beacon patterns that illuminate carefully chosen intervals of the angular space. Measurement of several such beacons can be used to identify the strongest paths in the channel. This method suffers from pattern imperfections caused by strong sidelobes that distort measurements. The sensing procedure is also disrupted by the possibility of destructive combination of paths that fall inside different bins in one beacon. 

Our approach in this paper is inspired by recent work on compressive phase retrieval \cite{bahmani2015efficient}, which cascades phase retrieval with coherent compressed sensing to reconstruct a sparse signal from noncoherent compressive projections.  The present paper differs from \cite{bahmani2015efficient}
in two key respects. First, rather than multiplying matrices known to be effective for phase retrieval and compressive sensing to obtain the measurement
matrix, we provide a virtual decomposition that enables use of a measurement matrix that can be realized with coarse phase-only control. Second, 
we are interested in continuous-valued (``off-grid'') parameter estimation rather than estimation of a signal that is sparse in a discrete basis, hence we replace the coherent compressive sensing stage by coherent compressive estimation. 

Compressive recovery of sparse signals with incomplete measurements has been studied in several previous works.
 In \cite{Heath_1bit, boufounos20081},
the problem of sparse signal recovery from 1-bit quantized projections is considered. In this case the sign of compressive projections is measured by the receiver and amplitude information is effectively lost. Such a problem can be formulated as a constrained
$\ell 1$ norm minimization and solved efficiently.
The loss of phase information, however, is more challenging.
Several interesting approaches to sparse phase retrieval have been proposed in recent literature, including \cite{schniter2015compressive,pedarsani2017phasecode,ohlsson2012cprl}.

All of these methods are designed to recover a sparse vector and, when applied to path recovery on the continuum of spatial frequencies, suffer from grid mismatch error. This error can only be reduced by oversampling the spatial frequency continuum, which results in a larger ``sparse" vector that requires more measurements to recover, effectively defeating the purpose of compressive estimation.

\begin{figure}[t]
\centering
\includegraphics[width=0.9\columnwidth]{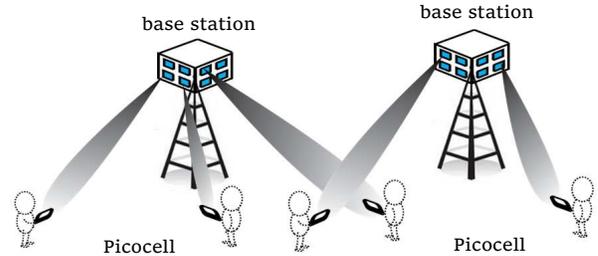}
\caption{Base station to mobile communication using narrow pencil beams in the dense picocellular network.}
\label{fig:picocells}
\end{figure}

\section{System model \label{sec:system_model}}

We consider the link between a directional transmitter using a linear phased array antenna of size $N$ (our approach applies directly to two-dimensional arrays, but we restrict attention
to linear arrays for simplicity of exposition) and one or more receivers. The wireless channel of each link consists of a number of paths from the transmitter to the receiver. 
The channel response is the sum of the responses of each of the paths on the array weighted by their respective complex amplitude.
 The array response of a path
at angle of departure $\theta$ is equal to:
\[
{\bf a}(\theta)= \left[e^{j1\frac{2\pi}{\lambda}d\sin\theta},e^{j2\frac{2\pi}{\lambda}d\sin\theta}, ..., e^{jN\frac{2\pi}{\lambda}d\sin\theta}\right]^T
\]
where $d$ is the inter-element spacing of the array and $\lambda$ is the wavelength.

Defining the {\it spatial frequency}
 $\omega=\frac{2\pi}{\lambda}d\sin\theta$, the array response of a path at angle $\theta$ can be described in terms of the corresponding spatial frequency $\omega$ by
\[
{\bf a}(\omega)=[e^{j1\omega},e^{j2\omega}, ..., e^{jN\omega}]^T
\]
In the remainder of this paper paths will be characterized by their spatial frequency instead of angle of departure.
The net channel response on the $N$ dimensional array is consequently equal to:
\begin{equation}
{\bf h} = \sum_{k=1}^K \alpha_k {\bf a}(\omega_k)
\label{eq:h}
\end{equation}
where $\alpha_k$ and $\omega_k$ are the complex amplitude and spatial frequency of the $k$'th path.

\begin{figure}[t]
\centering
\includegraphics[width=0.8\columnwidth]{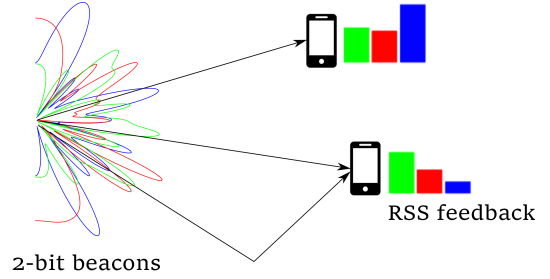}
\caption{Radiation pattern of compressive beacons on a 16-element array excited by weights from distribution $U(\{\pm 1, \pm j\})$, and feedback of RSS measurements made by mobile users.}
\label{fig:compressive_beacons}
\end{figure}

%
We assume that the transmitter broadcasts a series of $M$ beacons, and each receiver observes the strength of each beacon and provides an $M\times 1$ measurement vector as feedback to the transmitter at the end of the beaconing interval, which the transmitter employs to estimate the dominant spatial frequencies to the receiver. This procedure is depicted in Fig. \ref{fig:compressive_beacons}. 
Beacon $b$ excites the array with the randomly generated weight vector ${\bf w}_b=[w^b_1,w^b_2,...,w^b_N] ^T$ which sprays the emitted power differently in different directions. The response of beacon $b$ in the direction of
spatial frequency $\omega$ is denoted by:
\begin{equation}
f_b(\omega)={{\bf w}_b}^T  {\bf a}(\omega).
\label{eq:spatial_freq_beacon_response}
\end{equation}

The measurement made at the receiver will be a combination of the beacon response of all paths weighted by their corresponding complex path amplitude:
\begin{align}
y_b={\bf w}_b^T{\bf h} &= \sum_{k=1}^K \alpha_k {{\bf w}_b}^T  {\bf a}(\omega_k) \nonumber \\
& =\sum_{k=1}^K \alpha_k f_b(\omega_k)
\label{eq:coherent_measurement}
\end{align}
Therefore each angle of departure (AOD) will have its own  $M$-dimensional beacon response ``signature'' and the weighted sum of the signatures of the paths in the channel arrives at the receiver.
Based on the feedback from the receiver, the transmitter estimates the AOD of the strongest path(s) and beamforms in that direction for communication.
Although the transmitter will generally beamform toward the strongest path in the channel, maintaining a library of multiple strong paths is useful for rapid recovery from sudden blockage of the strongest path, as well as for using alternate paths to manage inter-cell interference.  Due to time-variant channel conditions and mobility of the receiver, channel estimation is repeated periodically to track the channel. This tracking takes place at a high enough rate to ensure the time between consecutive measurements is shorter than channel coherence time.

For simplicity, we assume omnidirectional reception, as in the early work on coherent compressive estimation \cite{ramasamy2012compressive}.  Including the effect of receive arrays in noncoherent compressive estimation is an important topic for future work.


In our proposed system, compressive beacons are used for channel discovery with certain limitations on array weights. First, we assume no control over the amplitudes of the sensing matrix and confine the system to unitary weights. Second, we assume limited control over element phases to simplify the front end hardware; to this end, we allow 2-bit phase control for the sensing matrix so that array phases are selected from the set of $\{\pm 1, \pm j\}$. These limitations are crucial for minimizing hardware complexity as array sizes grow.
 As shown in \cite{ramasamy2012compressive}, the degradation in beamforming performance caused by severe phase quantization is negligible as array size grows, meaning sacrificing phase granularity for array size can be an advantageous tradeoff.

Another important limitation is the loss of coherence from one beacon to the next. Due to oscillator design and dynamics, there is an unknown frequency offset between the local oscillators at the transmitter and receiver, that is constantly changing at an unknown rate. This random offset translates into a random offset in the phase of the channel measured by each beacon, corrupting the phase information of beacon measurements. For this reason, we assume RSS-only measurements are made at the receiver, effectively changing the coherent measurement model of (\ref{eq:coherent_measurement}) to noncoherent intensity measurements formulated by (\ref{eq:noncoherent_measurement}).
\begin{equation}
y_b=\left| {\bf w}_b^T {\bf h} \right| ^2
      =\left|\sum_{k=1}^K \alpha_k f_b(\omega_k)\right| ^2
\label{eq:noncoherent_measurement}
\end{equation}
The next section presents our proposed algorithm for compressive estimation of sparse channels under the nonlinear measurement model of (\ref{eq:noncoherent_measurement}).

\section{Noncoherent compressive estimation}

While several algorithms have been proposed for compressive estimation of sparse vectors, we focus on the Newtonized orthogonal matching pursuit method (NOMP) which is a robust, computationally efficient, and scalable estimation framework suited for sparse estimation on a continuum \cite{mamandipoor2015newtonized}.
 NOMP operates on the basis that the compressive projections of different spatial frequencies are almost orthogonal so strong paths are detected one by one via template matching of the compressive measurements against a  dictionary of beacon responses of spatial frequencies on an oversampled grid. The response of each detected frequency is subtracted from the measurements and Newton descent is used to fine tune the extracted paths on the continuum at each stage.
When phase information is unavailable, however, quasi-orthogonality of the paths is diminished and NOMP, or other standard compressive estimation techniques, cannot be applied. Therefore, we propose a hybrid scheme for noncoherent compressive estimation as follows.

\subsection{The noncoherent channel estimation algorithm \label{sec:algorithm}}

In order to maintain the benefits of NOMP and inspired by the approach put forth in \cite{bahmani2015efficient}, our method poses the estimation problem as a concatenation of phase retrieval and coherent compressive estimation via NOMP. The measurement matrix $A$ is defined as the product of a phase retrieval matrix, $A_{PR}$ of size $M\times M_{CS}$, and a compressive sensing matrix, $A_{CS}$ of size $M_{CS}\times N$:
\[
A=A_{PR}A_{CS}
\]
The measurement vector is then denoted as
\[ {\bf y}=|A h|^2 = |A_{PR}A_{CS}  {\bf h}|^2 \]
where the channel ${\bf h}$ is defined in (\ref{eq:h}).

We introduce the 
auxiliary
 measurement vector ${\bf y}_{CS}=A_{CS}{\bf h}$. This vector, if retrieved, effectively provides (up to an unknown, and irrelevant, complex gain) coherent compressive measurements of the target channel ${\bf h}$ by sensing matrix $A_{CS}$. The algorithm is thus divided into two stages: the complex auxiliary measurement vector ${\bf y}_{CS}$ is first estimated from RSS observations ${\bf y}=|A_{PR}{\bf y}_{CS}|^2$ using a phase retrieval algorithm, and the resulting estimate is subsequently used for compressive estimation of ${\bf h}$ by sensing matrix $A_{CS}$. The two stages of the algorithm are summarized as follows.

\noindent {\bf Stage 1: Phase retrieval.}
While several different algorithms have been proposed for phase retrieval, each with different specifications and conditions,  we consider the method proposed in \cite{bahmani2015efficient} for our baseline analysis, since it has provable performance bounds. This method is shown to obtain the target vector up to a constant phase ambiguity with high probability under the condition that elements of the $M\times N$ sensing matrix $A_{PR}$ are i.i.d. Gaussian and the number of RSS measurements, or $M$, scales as $n\log n$ for target vector size of $n$.

For evaluating the algorithm performance on simplified front ends with quantized beamforming phases, we use the more robust Wirtinger Flow algorithm \cite{candes2015phase} that utilizes gradient descent with careful initialization of the target vector to insure convergence to the global minimum. The reader is referred to these texts for a detailed description of the two methods.

%
%
%
%

\noindent {\bf Stage 2: Compressive estimation.}
The complex valued estimates obtained for ${\bf y}_{CS}$ are used in this step to solve for ${\bf h}$ the compressive estimation problem,
\[
{\bf y}_{CS}=A_{CS} {\bf h}
\]
in which ${\bf h}$ is a sparse combination of $K$ continuous-valued spatial frequencies as expressed in (\ref{eq:h}). For self-contained exposition, we briefly review the
NOMP algorithm that we use to sequentially extract the spatial frequency components.
Iteration $t$ consists of the following steps:

\begin{itemize}
\item Subtract the response of all paths extracted so far from the measurement vector to obtain the residual vector ${\bf y}_r$;
\[ {\bf y}_r={\bf y}-\sum_{k=1}^{t-1} \hat{\alpha}_k {\bf a}(\hat{\omega}_k) \]

\item  By matching the residual observation vector with the oversampled response dictionary, identify the strongest frequency in the remaining mixture and its corresponding complex amplitude on the oversampled grid of spatial frequencies, $\Omega$;
\[ \hat{\omega}_t=\argmax_{\omega\in \Omega} G_r(\omega),\]
\[G_r(\omega)=|{\bf f}(\omega)^H{\bf y}_r|^2/ \|{\bf f}(\omega)\|^2 \]
\[ \hat{\alpha}_t={\bf f}(\hat{\omega}_t)^H{\bf y}_r/\|{\bf f}(\hat{\omega}_t)\| \]
Where the notation ${\bf f}(\omega)$ represents the vector $[f_1(\omega),f_2(\omega),...,f_M(\omega)]$ of responses to the $M$ beacons as described in (\ref{eq:spatial_freq_beacon_response}).

\item  Use Newton refinement to fine-tune the identified frequency with respect to the residual. Add the resulting frequency and its corresponding complex amplitude to the set of estimated paths;

\item  Use cyclic Newton refinement steps on all paths extracted so far to fine-tune frequencies and amplitudes with respect to the measurement vector ${\bf y}$.
\end{itemize}
The algorithm proceeds until the residual energy is lower than a threshold, indicating all dominant paths have been extracted. The reader is referred to \cite{mamandipoor2015newtonized} for a detailed discussion of implementation and performance guarantees of the algorithm.

To ensure successful channel recovery of a $K$-sparse channel, the sensing matrix $A_{CS}$ must satisfy the restricted isometry property (RIP), i.e. provide sufficient separation between any two $K$-sparse vectors in the observation space. Random matrices have been shown to provide this condition with high probability under suitable circumstances.
In particular, it has been shown previously in \cite{candes2008restricted} that if the elements of the $M_{CS}\times N$ matrix $A_{CS}$ are generated from an i.i.d. Gaussian 
 distribution, then $A_{CS}$ will satisfy the above property with high probability when $M_{CS}$ (the number of compressive measurements) scales as $K\log N$. These predictions are useful guidelines in designing the algorithm parameters and creating the measurement matrices, as well as optimally allocating resources to the two stages of the process.


\subsection{Generating the  sensing matrices}

The compressive sensing and phase retrieval matrices, $A_{CS}$ and $A_{PR}$, must satisfy certain requirements to ensure the performance of each stage. At the same time, in order to be compatible with the hardware requirements described in Section \ref{sec:system_model},  the combined matrix $A=A_{PR}A_{CS}$,
must take values in the set $\{\pm 1,\pm j \}$. Generating the required matrices for hardware implementation and processing is therefore not trivial. In fact, identifying an $M\times M_{CS}$ and $M_{CS}\times N$ pair of matrices whose product, an $M\times N$ matrix, is in the quantized space is an overcomplete problem that cannot be solved exactly. We use the following procedure to minimize the distance between the product matrix and its quantized version, and treat this distance as a measurement error that can be compensated for by increasing the number of measurements and using stable phase retrieval and compressive sensing algorithms.   

As mentioned in Section \ref{sec:algorithm}, i.i.d. complex Gaussian matrices have the necessary projection characteristics for phase retrieval and compressive sensing and are a good choice for $A_{PR}$ and $A_{CS}$.
%
In the first step, the elements of $A$ are generated independently from a uniform distribution over the allowed values, i.e.,
\[A(i,j) \sim U(\{\pm 1,\pm j\}),\]
and $A_{PR}$ and $A_{CS}$ are then defined so that 1) their matrix product is equal to $A$, and 2) they are i.i.d. complex Gaussian (exactly for one of the matrices, and approximately for the other).

After producing the net matrix $A$, we independently generate $A_{CS
}$ from an i.i.d. complex Gaussian distribution:
\[A_{CS}(i,j) \sim CN(0,2 \sigma^2), \quad \sigma^2=\frac{1}{2N}   \]

The phase retrieval matrix is then defined as:

\begin{align}
& A_{PR}  
= A
{A_{CS}}^+ \nonumber \\
&{A_{CS}}^+  ={A_{CS}}^H \left( A_{CS}{ A_{CS}}^H \right)^{-1}  \label{eq:pinvAcs}
\end{align}
where $A_{CS}^+$ is the pseudoinverse of $A_{CS}$.
We now argue that the elements of $A_{PR}$ are also approximately i.i.d. with complex Gaussian distribution. Since the elements of $A_{CS}$ are i.i.d. and zero mean, its rows are approximately orthonormal, which motivates the following approximation:
\[
A_{CS}{A_{CS}}^H \approx I_{M_{CS}}.
\]
Substituting in (\ref{eq:pinvAcs}) we find that ${A_{CS}}^+$ is approximately equal to ${A_{CS}}^H$, so that the elements of the latter can be approximated as  i.i.d. complex Gaussian. Element $(i,j)$ of $A_{PR}$ is therefore equal to:
\[
A_{PR}(i,j) = \sum_{k=1}^{M_{CS}}{ A(i,k) {A_{CS}}^+ (k,j) } \approx \sum_{k=1}^{M_{CS}}{ A(i,k) {A_{CS}} (j,k) }
\]
It is easy to see that the elements of $A_{PR}$ are zero mean and uncorrelated.  Since each is the sum of a moderately large number of zero mean, independent random variables, we can invoke the central limit theorem to argue that the elements of $A_{PR}$ are jointly complex Gaussian, and therefore i.i.d. 

The product $A=A_{PR}\times A_{CS}$ is then reevaluated and quantized to obtain the final sensing matrix.


\begin{figure}[t]
\centering
\subfigure[K=2]
{\includegraphics[width=.9\columnwidth]{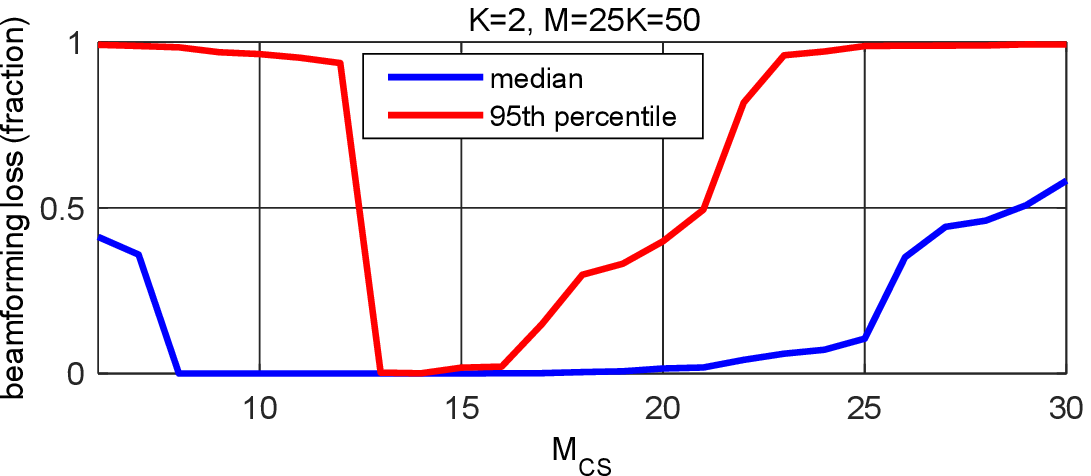}}
\subfigure[K=4]
{\includegraphics[width=.9\columnwidth]{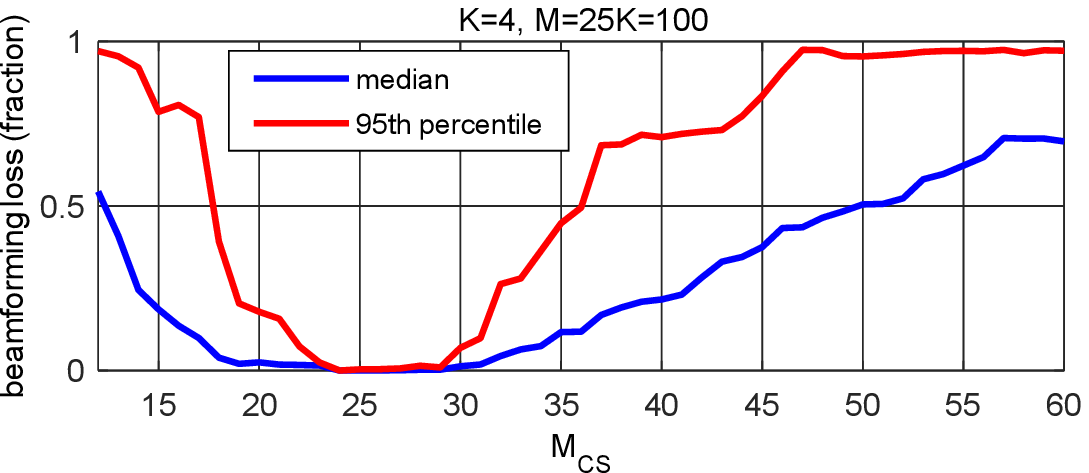}}
\caption{Beamforming error of estimated paths as a function of $M_{CS}$ for a fixed observation size $M$.}
\label{fig:optimal_Mcs}
\end{figure}

\begin{figure}[t]
\centering
\includegraphics[width=1\columnwidth]{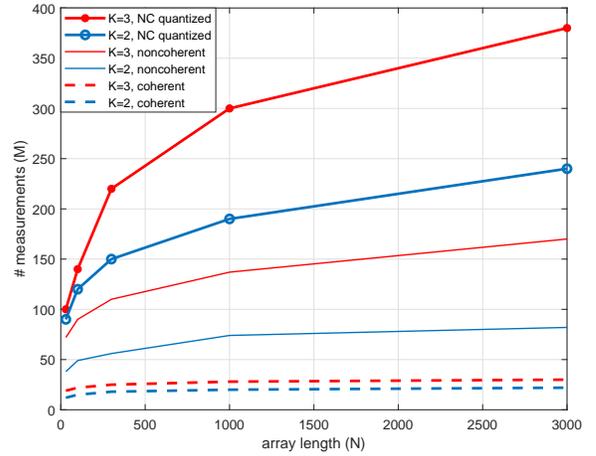}
\caption{Required number of measurements $M$ for 99\% probability of accurate channel recovery (beamforming loss $\leq$ 1dB).}
\label{fig:scaling}
\end{figure}

\section{Simulation results \label{sec:results}}

Both stages of the algorithm must succeed in order to accurately recover all strong paths in the channel. This requires appropriate choices for $M$ and $M_{CS}$.
The number of measurements required for coherent compressive sensing of an $N$-dimensional $K$-sparse vector is known to scale as $O(K\log N)$ \cite{candes2008restricted}.
The conditions for compressive estimation of a spatial channel with $K$ significant components are more complex \cite{ramasamy_TSP2014}, but when the components
are separated ``well enough," the number of measurements scales with $O(K\log N)$  as well (Theorem 4 in \cite{ramasamy_TSP2014}).
On the other hand, successful phase retrieval also poses restrictions on the relation between $M$ and $M_{CS}$. One fundamental lower bound for the number of observations required to identify $M_{CS}$ complex values is $M\geq 2M_{CS}$ since each complex variable is defined by two real values. The best algorithms available until now perform this estimation using $M=O(M_{CS}\log M_{CS})$ real-valued observations \cite{bahmani2015efficient}.

Since $M_{CS}$ scales as $K\log N$ and $M$ scales as $M_{CS}\log M_{CS}$ (with the current state of the art), we conclude that the number of measurements required for the proposed algorithm scales with array size $N$ and number of paths $K$ as:
\[
M=O(K\log N \log (K \log N)).
\]
This is only slightly more costly than the $O( K \log N )$ measurement complexity of coherent compressive estimation.
The proposed algorithm thus scales well with array size for sparse channels with a small number, $K$, of dominant components. Fig. \ref{fig:scaling} depicts simulation results for the required number of measurements for accurate channel recovery as a function of array size for different values of $K$. These results are in agreement with the preceding theoretical predictions.

For a sufficiently large number of measurements $M$,
the choice of $M_{CS}$ must provide an optimum tradeoff between phase retrieval and compressive sensing accuracy. This tradeoff is demonstrated in Fig. \ref{fig:optimal_Mcs}, where the overall performance of the algorithm is quantified for a fixed sensing configuration ($N=1000, M=25K, K=2,4$) as a function of $M_{CS}$.
The best choice for $M_{CS}$, given by the lowest point of the curve, corresponds to the tradeoff between the accuracy of phase retrieval and compressive sensing that yields the best beamforming performance.

\section{Conclusions}

We have shown that it is possible to estimate a sparse multipath channel with RSS measurements that can be realized using RF beamforming with
coarse phase-only control.  The approach inherits the desirable theoretical properties of phase retrieval and coherent compressive estimation, and 
is shown to be effective via simulations. In future work, we plan to validate the proposed approach on our mm wave testbed, and to explore techniques for reducing computational complexity of the phase retrieval component of our algorithm.  Another important topic for future investigation is to account for the receive antenna array in more detail.

\section*{Acknowledgments}
This work was supported by  NSF grants CNS-1317153 and CNS-1518812, and ComSenTer, one of six centers in JUMP, a Semiconductor Research Corporation (SRC) program sponsored by DARPA. 

\bibliographystyle{IEEEtran}
\bibliography{references}

\end{document}